# Use of a magnetic fluid for particle size analysis by a sedimentation method

Yury Dikansky, Arthur Zakinyan, Marita Bedganian

Department of Physics, Stavropol State University, 1 Pushkin Street, Stavropol 355009, Russia

**Abstract**

A new method of particle size analysis of micrometer-sized particles is discussed. The improved method of sedimentation analysis with magnetic fluids has the potential and versatility to characterize polydisperse systems.

**Keywords:** sedimentation analysis; magnetic fluid; microparticle size measurement; size distribution; powdery substances.

E-mail: zakinyan.a.r@mail.ru

The particle size measurement of a material is important in understanding its physical and chemical properties. In several industrial applications, knowledge of the product size distribution is critical for controlling and improving product quality and handling (e.g. in a food industry, pharmaceutical production, building industry, etc.). A sedimentation analysis is often applied for microparticle size measurements [1]. This method is based upon study of the terminal velocity acquired by particles sedimenting in a viscous liquid, i.e., larger particles sink faster than smaller particles when suspended in a liquid. In this paper, we report an improved method of sedimentation analysis of size distribution of microparticles of a powder and granular materials.

The basic difference of this method is the use of a magnetic fluid as medium, in which the motion (sedimentation) of particles is explored. A magnetic fluid (or ferrofluid) is a colloidal suspension of ultra-fine ferro- or ferri-magnetic nanoparticles suspended in a carrier fluid. If micrometer-sized non-ferromagnetic particles are added to a magnetic fluid, then the non-ferromagnetic inclusions form 'magnetic holes' in the fluid. It allows studying features of a motion of non-magnetic bodies in such media by measuring the change of magnetic properties at different points of the medium. In particular, the analysis of motion of particles and determination of their sizes can be carried out by measuring local changes of the magnetic permeability of the medium in which they move.

Figure 1 shows the particle measurement experimental setup, which was used for these purposes. The experimental setup consists of a vertical glass tube (length ~ 1 m, diameter ~ 0.01 m) filled with magnetic fluid. We used a kerosene-based magnetic fluid with dispersed magnetite nanoparticles of about 10 nm diameter stabilized with oleic acid. The magnetic fluid used was made by OJSC "NIPIgaspererabotka" (Russia). The magnetic fluid had a magnetic permeability $\mu \approx 7$. On the tube, the inductance coil of small length (5 mm) was located. The coil is connected to the measuring bridge (in our experiment "LCR meter-817"), which determines value of inductance of the coil. The magnetic measuring field produced by the coil is small enough (~ 50 A/m), so that its influence to any essential ad-hoc processes can be neglected.

A small quantity (~ 1 g) of the powder sample particles to be investigated is poured into the tube with the magnetic fluid. Timing with a stopwatch and measuring the inductance of the coil simultaneously starts. We can consider that all particles begin to move simultaneously and separately from each other. Thus, the particles are physically fractionated according to size before detection. When particles reach the place in the tube with magnetic fluid where the measuring coil is located, the inductance of the coil will change.

The sedimentation curve, $L(t)$, can be constructed by results of measuring the change of inductance, $L$, of the coil caused by the sedimentation of particles and the time, $t$, relevant to this change. As an example, the sedimentation curves, measured for particles of a diamond powder (curve 1) and a sand (curve 2), are shown in Fig. 2. The inductance was measured accurately to $0.01\,\mu H$, and the time was measured accurately to 0.1 s. Each powdered sample has been measured three times, and the obtained results are within 5% of each other.

The change of inductance must be proportional to the change of magnetic permeability of a magnetic fluid

$$\Delta L = k_1 \Delta \mu \qquad (1)$$

where $k_1$ is the constant of proportionality, $\mu$ is the macroscopic magnetic permeability of a magnetic fluid with microparticles.

For observance of a requirement of independence of a motion of individual particles the concentration of particles must be low ($\varphi \ll 1$). In this case, the change of a magnetic permeability in comparison with initial value for a pure magnetic fluid is proportional to the volume fraction of studied particles, as it follows from the Maxwell equation [2]

$$\Delta \mu = k_2 \varphi \qquad (2)$$

where $k_2$ is the constant of proportionality, $\varphi$ the particle volume fraction in the location of measuring coil. The optimal value of $\varphi$ is about 0.5%.

Let us present the whole population of measured particles as a set of monodisperse fractions. Then the particle volume fraction is given by

$$\varphi = V_i / V \qquad (3)$$

where $V_i$ is the volume of particles of some fraction, $V$ is some constant volume. Combining Eq. (1) with (2) and (3) we find

$$\Delta L_i = k V_i \qquad (4)$$

where $k = k_1 k_2 / V$, $\Delta L_i$ is the change of inductance, happening when particles of some fraction reach the place of the tube with a magnetic fluid, where the measuring coil is located. From Eq. (4) we can find each particle volume fraction or particle mass fraction:

$$\frac{V_i}{\sum V_i} = \frac{m_i}{m} = \frac{\Delta L_i}{\sum \Delta L_i} \qquad (5)$$

where $m_i$ is the mass of particles of some fraction, $m$ the total mass of particles in a sample. Using Eq. (5) we can construct the particle size distribution curve or histogram. We also can find a relative number of particles of some fraction, which is given by

$$\frac{N_i}{N} = \frac{\Delta L_i}{r_i^3} \bigg/ \sum \frac{\Delta L_i}{r_i^3} \tag{6}$$

where $r_i$ is the equivalent radius of particles of some fraction. The value of $r_i$ is given by the Stoke's equation

$$r_i = \sqrt{9\eta h / 2 g t_i (\rho_1 - \rho_2)} \tag{7}$$

where $\rho_{1,2}$ is the density of the particle and the medium, $g$ the acceleration of gravity, $\eta$ the dynamic viscosity of the medium (magnetic fluid), $t_i$ the time of the particle motion, $h$ the path length. We assume that the values of $\rho_1$, $\rho_2$, $\eta$ are known and the particles all have the same density. In our case, $h$ is a distance from the top of magnetic fluid in the tube to the position of measuring coil. Measured change of inductance of the coil, $\Delta L_i$, corresponds to the measurement time, $t_i$.

Thus, measuring the change of inductance of the coil and the time relevant to this change and using Eqs. (7) and (6) or (5), the particle size distribution of a powder may be obtained. As an example, the calculated particle size distribution curves of a diamond powder (curve 1) and a sand (curve 2), are shown in Fig. 3. These calculations are based on the sedimentation curves shown in Fig. 2.

The presented method allows to measure particle size distributions in the size range of 1–400 μm. Sub-micrometer particles cannot be reliably measured due to the effects of Brownian motion. Particles greater than about 400 μm cannot be measured because the time of acquiring the terminal velocity for these particles is too long. The method allows to measure weakly magnetic materials, and all particles of the sample should have the same density.

In several cases at late stages of sedimentation the analysis process can be sped up. For this purpose, the measuring coil is manually moved in small steps along the tube to scan and measure the finest particles of the sample.

It should be noted that the method of studying microparticles using magnetic fluids was previously offered in a publication [3]. But that method completely differs from method offered in this work and consists in use of an optical microscopy for study of a monolayer of microparticles immersed in a magnetic fluid under the action of magnetic field.

The procedure presented in this paper is easier then classical sedimentation analysis. The important feature of an offered method in comparison with a traditional sedimentation analysis is lack of the extremely inexact operations of double differentiation. This method combines the simplicity with the ability to measure not only the average particle size, but also the detailed particle size distribution. The technique of using magnetic fluids for measuring the size distribution of the particles could be useful in several areas of study such as soil science or the life sciences.

The authors have taken out a patent for this invention [4].

**Acknowledgment:** This work was supported by the Federal Education Agency of the Russian Federation in Scientific Program "Development of Scientific Potential of Higher School".

**Figure Captions**

Fig. 1. Experimental setup for particle size measurements.

Fig. 2. Sedimentation curves: 1 – diamond powder, 2 – sand.

Fig. 3. Particle size distribution curves: 1 – diamond powder, 2 – sand. Here $\Delta r_i$ is the difference of radii of particles of adjacent fractions.

Figure 1. Experimental setup for particle size measurements.

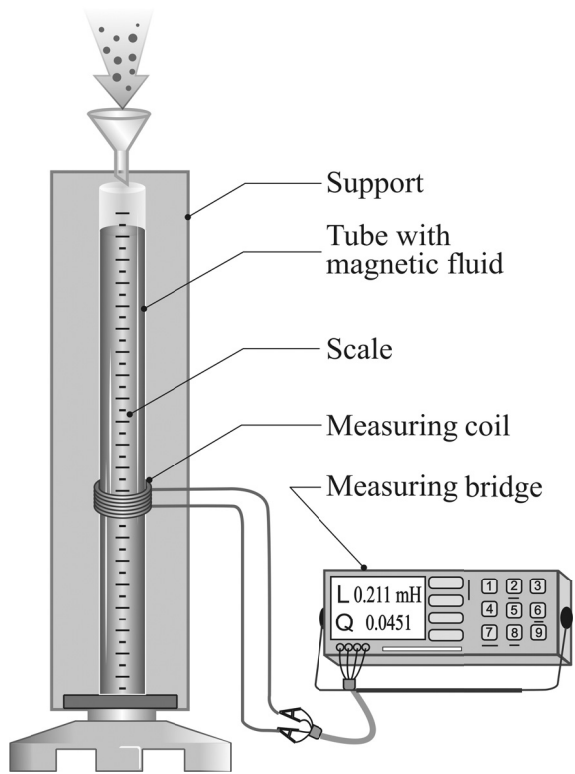

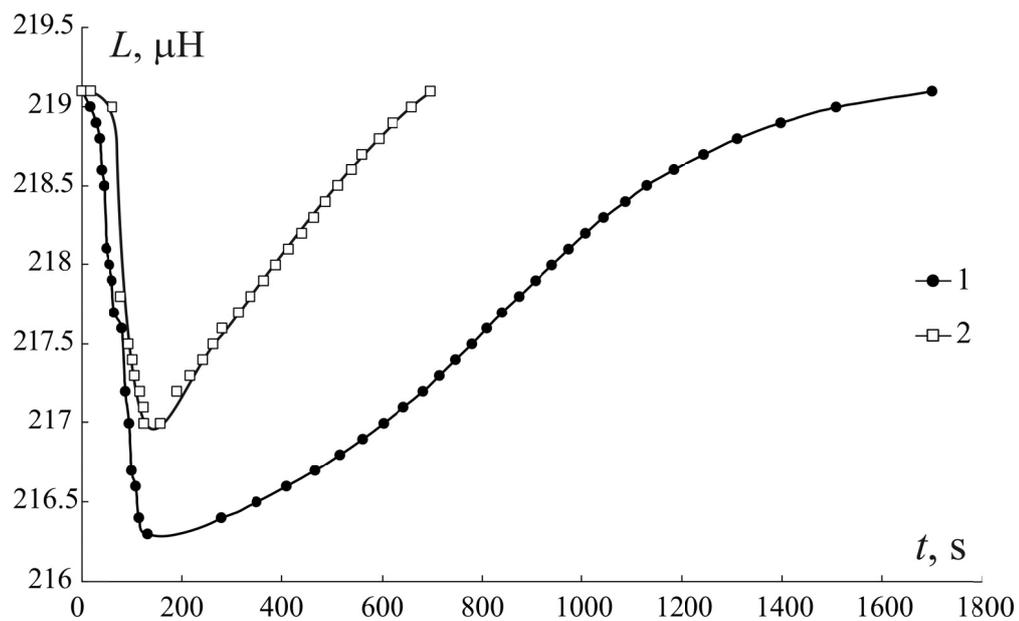

Figure 2. Sedimentation curves: 1 – diamond powder, 2 – sand.

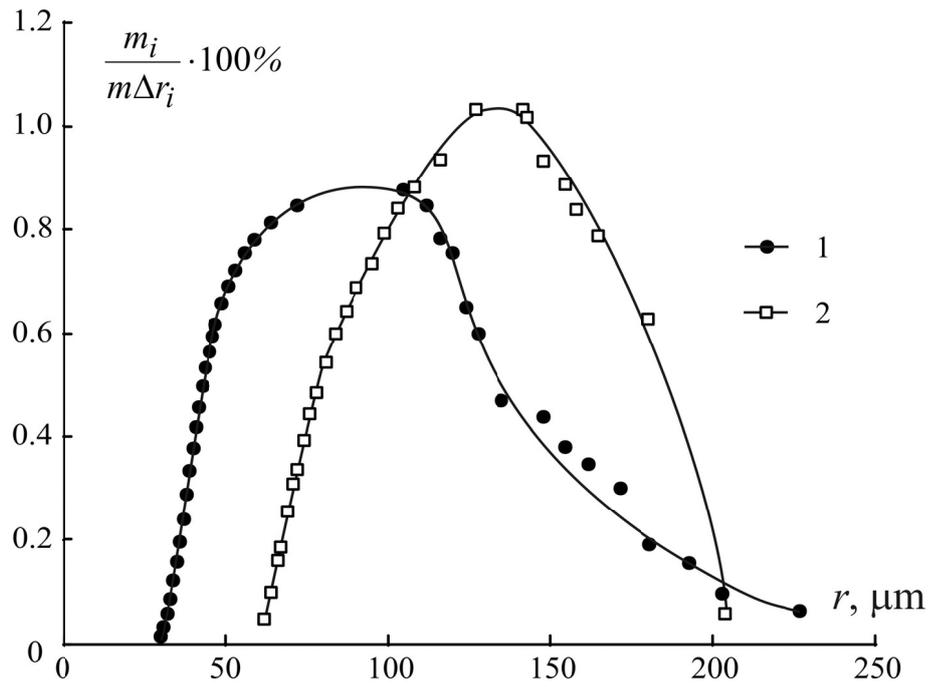

Figure 3. Particle size distribution curves: 1 – diamond powder, 2 – sand. Here $\Delta r_i$ is the difference of radii of particles of adjacent fractions.